\documentclass[aps,pre,twocolumn,float]{revtex4}
\usepackage{amsmath,bm,epsfig}

\let\*\cdot
\def\<{\left\langle} \def\>{\right\rangle} \def\({\left(} \def\){\right)}
 \let\~\widetilde \let\^\widehat 

\def\be{\begin{equation}}\def\ee{\end{equation}}
\def\bea{\begin{eqnarray}}\def\eea{\end{eqnarray}}
\def\bse{\begin{subequations}}\def\ese{\end{subequations}}
\newcommand{\BE}[1]{\begin{equation}\label{#1}}
\newcommand{\BEA}[1]{\begin{eqnarray}\label{#1}}
\newcommand{\BSE}[1]{\begin{subequations}\label{#1}}

\usepackage{pstricks}
\usepackage{pst-node}
\usepackage[ansinew]{inputenc}
\usepackage{amssymb,amsmath}

\def\BSE{\begin{subequations}}\def\ESE{\end{subequations}}
\let \= \equiv

\def\b{\beta}

\def\o{\omega}

\def\be{\begin{equation}}       \def\ba{\begin{array}}

\def\ee{\end{equation}}         \def\ea{\end{array}}

\def\bea {\begin{eqnarray}}      \def\eea {\end{eqnarray}}

\def\bean{\begin{eqnarray*}}    \def\eean{\end{eqnarray*}}

\def\e{\varepsilon}

\def\<{\langle} \def\({\left(}  \def\>{\rangle} \def\){\right)}

\newtheorem{exi}{Example}

\begin{document}

\title{Final comment to the results of Bustamante et al. on discrete Rossby/drift wave resonant and quasi-resonant triads}
\author{A. Kartashov, E. Kartashova}
   \begin{abstract}
In this final note we demonstrate that the authors of manuscripts arXiv:1210.2036, arXiv:1309.0405 and arXiv:1309.5513 use mathematical notations and notions sometimes in the
 standard meaning and sometimes in a sense which differs from the standard. As this specific use is not  defined beforehand, the authors' statements are self-contradictory which makes any further scientific discussion meaningless.
\end{abstract}


\maketitle
\section{Introduction}
The language of mathematics has been developed for helping to avoid  ambiguities existing in plain colloquial language; each notion and notation in mathematics is strictly defined which makes it possible, for instance,  to translate the following paragraph

\emph{"Rule to solve $x^3 + mx = n.$
Cube one-third the coefficient of x; add to it the square of one-half the constant of the equation; and take the square root of the whole. You will duplicate this, and to one of the two you add one-half the number you have already squared and from the other you subtract one-half the same... Then, subtracting the cube root of the first from the cube root of the second, the remainder which is left is the value of x (Gerolamo Cardano, Ars Magna, 1545). "}

into a nice simple formula

\be
\sqrt[3]{\frac{n}{2}+\sqrt{\frac{n^2}{4}+\frac{m^3}{27}}} -\sqrt[3]{-\frac{n}{2}+\sqrt{\frac{n^2}{4}+\frac{m^3}{27}}}
\ee

This "translation" is  possible because all notions and notations used ("equation", "solution", "square root","one-half", etc. etc.) are strictly defined and all mathematicians use them in the same sense. Any mathematical discussion about the properties of $\sqrt{4}$ would be meaningless if for  one disputant $\sqrt{4}= \pm 2$, and for the other  $\sqrt{4}=-13/3$ .

Below we give some examples of misuse of various mathematical notions found in [1]-[3].

\section{Notions: "smaller" and "smaller or equal"}
It is stated in [1], p. 3, that the authors of [3] "confined their search to the interior of the domain $0\le |k_x|,|k_y| \le 200$" thus trying to explain why some solutions have been lost. Correct mathematical notation in this case would be $0<|k_x|,|k_y| < 200$.

Obviously, the authors have great problems with the notions of "smaller" and "larger" anyway: in [2] they write:

\emph{The success of [3] is that the observed
features (connectivity and percolation) for such small sample have the same properties as the
corresponding features for the whole set of triads, computed directly by brute force in a thorough
study at higher resolution ($L \ge 256$) published by Bustamante and coworkers in [4].}

However, in [4] only solutions are studied which  have wave numbers less or equal to 256, i.e. $L \le 256$ (see Fig.5, [4]).

One can only wonder what "the success of [3] is".

\section{Notion: "limit"}
To say that
\be \label{limit}
    \lim_{x \to p}f(x) = L, \,
\ee
means that $ƒ(x)$ can be made as close as desired to $L$ by making $x$ close enough - but not equal - to $p$.  This is the standard $\e-\delta$ definition of a limit, which can be found in any college course of mathematical analysis.

In [5], the Theorem 3 has been proved for the case $\b \to \infty$ which means for any BIG ENOUGH BUT FINITE magnitude of $\b$.

It is stated in [2], p.2, that the results of [5] are "valid in the asymptotic limit of infinitely large $\b$, and can not be used "to deduce results for finite $\b$". This statement is in contradiction with the standard definition of a limit.

By the way, the notion of "asymptotic limit" is not even used in [5].

It is also stated in  [2], p.2, that the results of [3] are obtained "for small or moderate $\b$". We
 were unable to confirm this statement with a quote from [3].

 Symbol $\b$ can be found in [3]  altogether 4 times: "so-called $\b$-plane approximation" (p.3), in the formulae (1), (3) and the formula without number (at the beginning of Sec. 6.). Not a word is written about magnitude of $\b$ - whether it is small or big or infinite.

In accordance with standard mathematical conventions this means that \emph{the results}  [3] \emph{are formulated for arbitrary magnitude of $\b$}, and in particular they should also be valid for big $\b$.
 \section{Notion: "solution"}
It is stated in [1], p. 1, that "Fourier wavevectors ... can interact if and only if"  resonance conditions (\ref{res-vect}),(\ref{res-dis}) shown below, are satisfied:
\bea
k_1+k_2=k_3, \, l_1+l_2=l_3, \label{res-vect}\\
\o_1+\o_2=\o_3 \label{res-dis}
\eea

 "The set of ... wavevectors satisfying (\ref{res-vect}),(\ref{res-dis}) is called resonant set" ([1], p. 1).  According to this definition

\emph{any set of 3 wavevectors, satisfying (\ref{res-vect}),(\ref{res-dis}), is A SOLUTION} (called exact resonant triad).

Again, according to the given definition,  two different sets of wavevectors, e.g. $\{(-18,-46),(2,-22),(-16,-68)\}$ and $\{(-18,-46),(16,68),(-2,22)\}$, are TWO SOLUTIONS, not one. The fact that these two solutions can be transformed into each other by a simple algebraic operation does not make them one solution.

Consequently, \emph{all six modes have to be counted in the complete set of resonant modes} and not only one of them as it is done in [3].

 The authors of [1] argue that as e.g. solutions $\mathbf{k}_1,\mathbf{k}_2,\mathbf{k}_3$ and $-\mathbf{k}_1,-\mathbf{k}_2,-\mathbf{k}_3$ have  dynamical equations for the amplitudes differing only in sign, they should be regarded as ONE solution of (\ref{res-vect}),(\ref{res-dis}).

This argument has two flaws (one coming from kinematics and another from dynamics) which make it meaningless in the context of discussion of the results presented in [3]:

$\diamond$ \emph{\textbf{Kinematics}}.\\
The definition of resonant modes/triads given in [3] DOES NOT INCLUDE any dynamical characteristics such as coupling coefficients or aspecial form of dynamic equations or whatever. This information was introduced - by extended group of authors, [1] - AFTER the publication of [3], in another manuscript, and is irrelevant for the  discussion of [3].  Omission of these important dynamical characteristics would be pardonable for young beginners  in scientific work [3] -  in the case if only EXACT RESONANCES were studied.

$\diamond$ \emph{\textbf{Dynamics}}.\\
However, as quasi-resonances were studied in [3] also, this omission has led to a completely false general picture - both resonant and quasi-resonant modes are calculated incorrectly. This can be seen immediately from the Fig.4, [3] where the number of quasi-resonant triads is represented by NONSYMMETRIC HISTOGRAM - and it is completely obvious that the diagram MUST BE symmetric. Namely,
as a quasi-resonance is formed by modes satisfying
\bea
k_1+k_2=k_3, \, l_1+l_2=l_3, \label{q-res-vect}\\
|\o_1+\o_2-\o_3| \le \delta, \label{q-res-dis}
\eea
this means in particular that any solutions $\mathbf{k}_1,\mathbf{k}_2,\mathbf{k}_3$ and $-\mathbf{k}_1,-\mathbf{k}_2,-\mathbf{k}_3$ would satisfy (\ref{q-res-vect}), (\ref{q-res-dis}) with $\delta$ and $-\delta$ detuning correspondingly. Moreover, in this case, \\
DYNAMICAL EQUATIONS FOR SYMMETRICAL MODES WOULD BE DIFFERENT\\
and the argument fails completely.

It is interesting to notice that the authors say "in fact every physically sensible triad is quasi-resonant" ([3], p.10) thus making  their own arguments null and void.

\section{Summary}
The results of [3] can be briefly presented as follows:

$\diamond$ \emph{\textbf{Kinematics}}: A method is presented which allows to compute SOME exact resonance triads. No estimation is given about what part of the all exact resonant triads has been found.

$\diamond$ \emph{\textbf{Kinematics}}: A method is presented which allows to compute SOME quasi-resonance triads. The method is statistically biased; the best quasi-resonance found by this method has detuning  over six orders decimal magnitude worse than the really best in the domain $L \le 200$: $\sim 10^{-5}$ versus $\sim 8.95 \cdot 10^{-12}$.

$\diamond$  \emph{\textbf{Dynamics}}: Any study of dynamics is omitted  which yields a completely erroneous picture of energy percolation through the spectrum, basing only on exact and quasi-resonance triads available in [3].

\section{Acknowledgement}
As we first read [3] and established the facts formulated in the previous section, we sent a few emails to M. Bustamante willing to discuss them. As M. Bustamante informed us that  he has no time for a discussion, we have decided to study the problem ourselves in [6].

Now that the debate went public, we hope that our findings would be of help for  other researchers working in the area of discrete resonances and quasi-resonances of Rossby/drift waves, e.g. [4],[7],[8], etc.

\emph{ArXiv} publications  give researchers a  unique opportunity to really discuss various physical and mathematical problems occurring during the research - it is sufficient  to recall a perennial discussion ventured in arXiv by V. Lvov and S. Nazarenko about the energy cascade in  super fluids, see e.g. \emph{arXiv:0909.2936},  \emph{arXiv:1208.4593}.

In the name of all researchers we are deeply grateful to the creators and maintainers of the internet portal \emph{arXiv} for their valuable job.

\textbf{{References.}}

[1] M. Bustamante, U. Hayat, P. Lynch, B. Quinn.
\emph{arXiv:1309.0405}

[2] M. Bustamante, U. Hayat, P. Lynch, B. Quinn.
\emph{arXiv:1309.5513}

[3] M. Bustamante, U. Hayat.
\emph{arXiv:1210.2036}

[4] J. Harris, C. Connaughton, M. Bustamante.
\emph{arxiv:1212.5336}

[5] M. Yamada, T. Yoneda.
\emph{Physica D} \textbf{245} (2013): 1-7.

[6] A. Kartashov, E. Kartashova. \emph{arXiv:1307.8272}

[7] M. Bustamante, B. Quinn.  \emph{arXiv:1305.5517}

[8] K. Harper, M. Bustamante, S. Nazarenko.
\emph{arXiv:1212.3156}
\end{document}